\documentclass[fleqn, usenatbib, useAMS]{mnras}

\usepackage[T1]{fontenc}
\usepackage{upquote} 
\usepackage[british]{babel}
\usepackage{epsfig}
\usepackage{color}
\usepackage{amsmath} 

\newcommand{\Msolar}{\rmn{M}_{\odot}} 
\newcommand{\Rsolar}{\rmn{R}_{\odot}} 
\newcommand{\Lsolar}{\rmn{L}_{\odot}} 
\newcommand{\MWD}{M_{\rmn{WD}}}
\newcommand{\MMS}{M_{\rmn{MS}}}

\addto\captionsbritish{%
}

\title[HeWD+MS mergers: single low-mass WDs]{Evolution models of helium white dwarf--main-sequence star merger remnants: the mass distribution of single low-mass white dwarfs}
\author[X.~Zhang et al.]{Xianfei Zhang$^{1}$\thanks{E-mail: zxf@bnu.edu.cn},
Philip D.~Hall$^{2}$,
C.~Simon Jeffery$^{2,\,3}$,
and Shaolan Bi$^{1}$\\
$^1$Department of Astronomy, Beijing Normal University, Beijing, 100875, China\\
$^2$Armagh Observatory and Planetarium, College Hill, Armagh BT61 9DG, UK\\
$^3$School of Physics, Trinity College Dublin, Dublin 2, Ireland}

\date{Accepted XXX. Received YYY; in original form ZZZ}
\pubyear{2017}

\begin{document}
\label{firstpage}
\pagerange{\pageref{firstpage}--\pageref{lastpage}}
\maketitle

\begin{abstract}
  It is not known how single white dwarfs with masses less than $0.5\,\Msolar$ -- low-mass white dwarfs -- are formed.
  One way in which such a white dwarf might be formed is after the merger of a helium-core white dwarf with a main-sequence star that produces a red giant branch star and fails to ignite helium.
We use a stellar-evolution code to compute models of the remnants of these mergers and find a relation between the pre-merger masses and the final white dwarf mass.
Combining our results with a model population, we predict that the mass distribution of single low-mass white dwarfs formed through this channel spans the range $0.37$ to $0.5\,\Msolar$ and peaks between $0.45$ and $0.46\,\Msolar$.
Helium white dwarf--main-sequence star mergers can also lead to the formation of single helium white dwarfs with masses up to $0.51\,\Msolar$.
In our model the Galactic formation rate of single low-mass white dwarfs through this channel is about $8.7 \times 10^{-3}\,\rmn{yr}^{-1}$.
Comparing our models with observations, we find that the majority of single low-mass white dwarfs ($\le 0.5\,\Msolar$) are formed from helium white dwarf--main-sequence star mergers, at a rate which is about $2$ per cent of the total white dwarf formation rate.
\end{abstract}


\section{Introduction}
White dwarfs (WDs) are the end state of most stars.
They can be classified by the dominant compositions of their cores: helium, carbon and oxygen (CO) or oxygen, neon and magnesium (ONeMg).
Most WDs with masses below $0.5\,\Msolar$ -- low-mass WDs -- are helium-core white dwarfs (HeWDs).
Most of these are found in close binaries with main-sequence star or WD companions \citep{Maxted99,Maxted2000,Zorotovic2011}.
The binary fraction depends on the mass of the WD: all of those with $M_{\rm WD} \le 0.25\,\Msolar$ (extremely low-mass WDs) have companions and about $20$--$30$ per cent of those with $0.25 < M_{\rm WD}/\Msolar \le 0.5$ do not have companions \citep{Brown11}.
Those with companions may have formed through phases of Roche lobe overflow or common-envelope evolution that stripped the envelopes of the progenitors of the WDs, red giant branch (RGB) stars \citep{Kippenhahn1967}.
The WDs without companions are more difficult to explain.
They are not thought to have formed through normal single stellar evolution because to do so would require a star of low zero-age main-sequence mass and a time longer than the age of the Universe \citep{Bergeron1992}.

Several channels to the formation of single low-mass WDs have been proposed.
(1) Single star evolution with enhanced mass-loss.
A star evolves with a greater than usual mass-loss rate and enough of the envelope is stripped for the star to contract when it is an RGB star and become a HeWD \citep{Castellani1993,dcruz96}.
(2) Mass ejection by massive planets.
A massive planet enters the envelope of an RGB star.
In the subsequent common-envelope phase the envelope is ejected, the planet evaporates and the stripped RGB star becomes a HeWD \citep{Nelemans98}.
(3) Supernova stripping.
An RGB star has most of its envelope stripped by a companion that explodes in a supernova.
The stripped RGB star becomes a HeWD \citep{Justham2009,Wang2009,Geier2015}.
(4) HeWD+HeWD merger.
Two close HeWDs merge as a result of gravitational-wave radiation.
If the total mass is small then He ignition is avoided and the remnant becomes a HeWD \citep{Saio00,Han02,Han03,Zhang12a}.
(5) HeWD+MS merger.
A main-sequence star (MS) and a HeWD merge to form an RGB-like star.
The RGB-like remnant evolves like a very low-mass star to become a HeWD \citep{Clausen2011,Zhang2017,Zorotovic2017}.
We focus on this last channel.

There are many short-period detached binary systems composed of HeWDs with MS companions that may be the immediate progenitors of HeWD+MS mergers \citep{Zorotovic2011}.
The formation rate of single low-mass WDs formed through these mergers depends on the stability of mass transfer when the orbital separation decreases such that the MS star begins to fill its Roche lobe.
If the MS is of low mass, $\MMS \le 0.7\,\Msolar$, it has a substantial convective envelope and mass transfer is traditionally expected to be dynamically unstable if $\MMS / \MWD > 0.695$ \citep{Hurley02,Shen2009} and lead to a merger.
\citet{Nelemans2016} suggested that nova common envelopes at the onset of mass transfer causes systems with $\MMS / \MWD < 0.695$ to merge.
Motivated to explain the lack of CVs containing HeWD accretors, \citet{Schreiber2016} suggested an equation for the angular momentum loss during mass transfer, consequential angular momentum loss (CAML), such that most HeWD+MS detached systems merge on becoming semidetached.
\citet{Zorotovic2017} considered the effect of this on the formation rate and mass distributions of single WDs formed through HeWD+MS mergers.
However, their work was based on simple assumptions about the evolution of merger remnants.
In this paper, we build on our work on the modelling of single hot subdwarfs as HeWD+MS merger remnants \citep{Zhang2017} to compute the masses of single low-mass WDs formed through the HeWD+MS merger channel.
We predict the mass distribution of single low-mass WDs and compare with observations.

\section{Method}
We use the stellar evolution code \texttt{MESA} \citep[Modules for Experiments in Stellar Astrophysics v8118;][]{paxton11,paxton13,paxton15}.
To make model merger remnants, we rapidly accrete H-rich matter onto HeWDs.
The initial HeWD models are made by stripping mass from a $1.5\,\Msolar$ RGB star evolved from the zero-age main sequence with metallicity $Z=0.02$.
We strip mass from the star when it evolves to the RGB phase and has a He core mass of $0.250$, $0.275$, $0.300$, $0.325$, $0.350$, $0.375$ or $0.400\,\Msolar$.
At each of these stages, a large mass-loss rate is switched on until the H envelope is completely stripped.
We take our starting model to be when the remaining He core evolves to the WD cooling track and a low logarithmic surface luminosity $\log (L/\Lsolar) = -2$ \citep{Zhang12a}.
In this way we produce models of HeWDs of the required masses which are used in subsequent steps.
We represent the merger by accreting H-rich matter at a rate of $1\,\Msolar\,\rm{yr}^{-1}$ on to the HeWD.
To match the composition of the MS star, the accreted material has He mass fraction $Y=0.28$ and $Z=0.02$ with the scaled metal mixture of \citet{Grevesse1998}.

Our chosen \texttt{MESA} input parameters are given in Appendix~\ref{sec:mesa_inlist}.
The mixing-length theory parameter $\alpha_{\rmn{MLT}} = 1.9179$ to match the solar calibration of \citet{paxton11}.
The opacity is taken from the OPAL Type 2 tables of \citet{Iglesias1996} and \citet{Ferguson2005} at low temperatures.
The outer boundary conditions are matched to an Eddington grey photosphere.
The post-merger models evolve with mass loss when the effective temperature is below $10^4\,\rm{K}$ and mass is lost according to Reimers' formula with $\eta_{\rm R} = 0.5$.
Mixing in the models is only by convection in convective regions and atomic diffusion in radiative regions.

\section{Merger remnants}
We model the remnants of mergers of different combinations of $\MWD$ and $\MMS$.
The remnant of a HeWD+MS merger has an RGB structure but with a more degenerate He core than is found in RGB stars that evolved from main-sequence stars.
As in an RGB star, H burns in a shell and the He produced is added to the He core, compressed and heated.
Wind is lost according to Reimers' mass-loss-rate formula with $\eta=0.5$ during this phase.

\begin{figure}
\centering
\includegraphics[width=84mm]{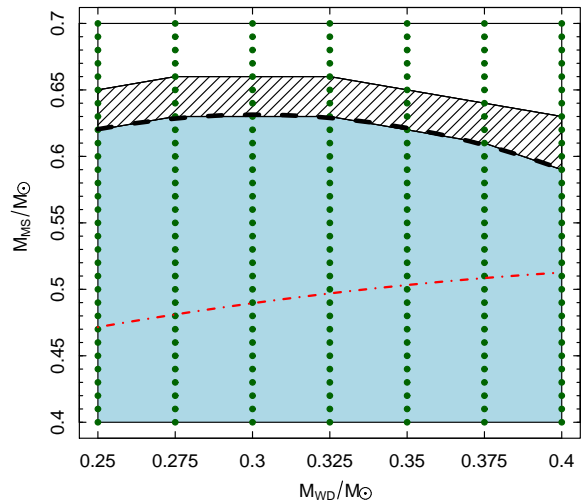}
\caption[]{HeWD+MS merger outcomes as a function of $\MWD$ and $\MMS$, when $\eta=0.5$ in the post-merger phase.
    All systems in the region shown merge according to the CAML hypothesis of \citet{Schreiber2016}.
  The models computed are indicated by dots.
  In the blue filled region, remnants become single HeWDs.
  In the hatched band, remnants become hot subdwarfs and then low-mass CO WDs.
  In the white region, remnants ultimately become CO WDs.
  The boundary between the formation of single HeWDs and hot subdwarfs is indicated by the thick dashed line.
  The dot--dashed line shows the maximum HeWD mass that can be formed from a HeWD+MS merger for a given pre-merger HeWD mass.}
\label{models}
\end{figure}

We have previously summarized the paths that are available to HeWD+MS merger remnants \citep{Zhang2017}.
Fig.~\ref{models} shows the combinations of $\MMS$ and $\MWD$ that result in the different paths.
In this work we focus on those that avoid He ignition and become HeWDs.
In Fig.~\ref{models} the thick dashed line,
\begin{equation}
  \MMS = -4.1905 \times \MWD^2 + 2.5238 \times \MWD + 0.2514,
\end{equation}
gives the upper boundary of this region -- the maximum $\MMS$ for a given $\MWD$, if the remnant is to avoid He ignition and become a single HeWD.
In this case the final mass of the WD depends on both $\MMS$ and $\MWD$.
An approximate fit is:
\begin{equation}
  M_{\rm f} = 0.3148 \times \MWD + 0.2048 \times \MMS + 0.2658,
\end{equation}
where $M_{\rm f}$ is the final mass of the WD, $\MWD$ is the mass of the WD and $\MMS$ is the mass of the MS star.
The maximum HeWD mass that can be formed from a HeWD+MS merger for a given pre-merger HeWD mass is also shown by the dot--dashed line in Fig.~\ref{models}.

The combinations of $\MWD$ and $\MMS$ that lead to the formation of a HeWD and the mass of the WD produced depend on the assumed $\eta$.
Estimates of $\eta$ from observations of normal stars suggest that it might range between about $0.1$ \citep{Miglio2012} and $0.7$ \citep{McDonald2015}, so as well as the standard $\eta=0.5$, we also compute models with $\eta=0.1$ and $0.7$.
If $\eta=0.1$, the maximum $\MMS$ is decreased to about $0.3\Msolar$ and the relation between $M_{\rmn{WD}}$ and $M_{\rmn{f}}$ becomes
\begin{equation}
M_{\rm f}=0.62 \times \MWD + 0.5 \times \MMS + 0.146.
\end{equation}
If $\eta=0.7$, then
\begin{equation}
M_{\rm f}=0.3587 \times \MWD + 0.2054 \times \MMS + 0.2265.
\end{equation}

\begin{figure}
\centering
\includegraphics[width=84mm]{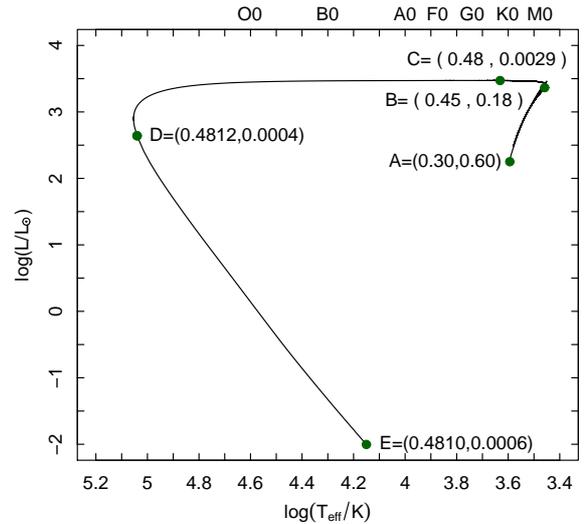}
\caption[]{Evolution in the theoretical Hertzsprung--Russell diagram after a $0.3$+$0.6\,\Msolar$ HeWD+MS merger.
  The core and envelope masses are indicated in the brackets at various stages, A to E.}
\label{hr30}
\end{figure}

Fig.~\ref{hr30} shows an example of the evolution in the theoretical Hertzsprung--Russell diagram of the remnant of an $\MWD+\MMS=0.3+0.6\,\Msolar$ merger that forms a single HeWD.
The core and envelope masses as the star evolves are indicated at various stages in the figure.
The star loses nearly half of its mass through Reimers' mass-loss on the RGB.
The HeWD produced is massive, about $0.48\,\Msolar$.
Our models show that HeWDs up to a mass of about $0.51\,\Msolar$ can be formed by HeWD+MS mergers.
This is a larger mass HeWD than could normally be formed in standard or binary evolution of a $Z=0.02$ star.
It takes about $1.5\times10^7 \rm yr$ to become a WD after the merger.

Mergers of HeWDs and MS stars may also result in the formation of single COWDs, either of relatively low mass for a CO-core WD ($\MWD \le 0.5\,\Msolar$) after a phase as a hot subdwarf, or after normal evolution through core He-burning and asymptotic giant-branch phases.
In Fig.~\ref{models}, these remnants arise for parameters above the thick dashed line.

\section{Population synthesis}
To compute the mass distribution of single low-mass WDs formed through the HeWD+MS merger channel, we compute the properties of a synthetic population of primordial binary systems.

For the joint distribution of zero-age parameters, the masses are generated according to the formula of \citet{Eggleton89} and the initial mass function of \citet{Miller79}, with masses in the range $0.08$--$100\,\Msolar$.
The distribution of orbital separations, $p(a)$, is that of \citet{Han98}:
\begin{equation}
  p(a) =
  \begin{cases}
    0.070(a/a_0)^{1.2}  & \phantom{a_0 \le} a \leq a_0 \\
    0.070              & a_0 \le a \le a_1, \\
  \end{cases}
\end{equation}
where $a_0=10\,\Rsolar$, $a_1=5.75 \times 10^6\,\Rsolar = 0.13\,\rm{pc}$.
According to this distribution of orbital separations, approximately $50$ per cent of stellar systems have orbital periods greater than $100\,\rmn{yr}$ \citep{Han98}; these systems are considered to be single stars.
We model a population with an age of $13\,\rm{Gyr}$ and a constant star formation rate history of $5\,\Msolar\,\rm{yr}^{-1}$, intended to represent the Galaxy \citep{yungelson98}.

To calculate the properties of a population with the chosen distribution of zero-age parameters and ages, we draw $10^7$ sets from the distribution of zero-age parameters.
We then use a rapid binary evolution code (\texttt{BSE}, \citealt{Hurley00,Hurley02}) to evolve these binary systems for $13\,\rm{Gyr}$.
The \texttt{BSE} input parameters in this work are the same as those used in previous work on modelling the Galactic rate of double WD mergers \citep{Han98, zhang2014}, except that we modify the code so that HeWD+MS stars merge when they have $(\MWD, \MMS)$ in the region indicated by the inclusion of CAML described by \citet{Schreiber2016} and \citet{Zorotovic2017}.

\begin{figure}
\centering
\includegraphics[width=84mm]{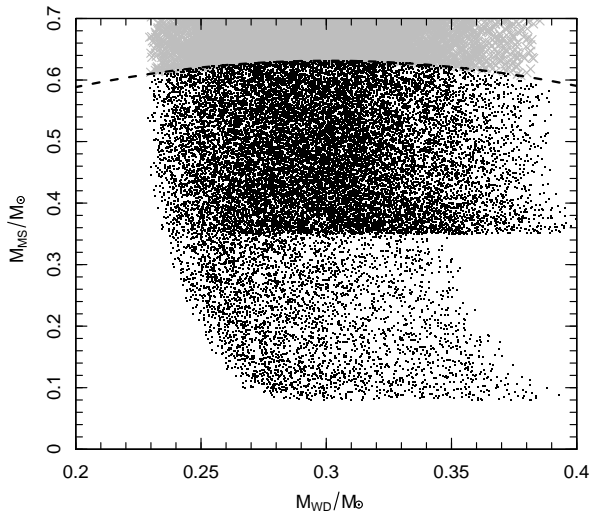}
\caption[]{Combinations of WD and MS mass, $\MWD$ and $\MMS$, at the onset of Roche-lobe filling realised at any age in our set of $10^7$ binary systems.
  All systems in the region shown merge according to the CAML hypothesis of \cite{Schreiber2016}.
  Below the dashed line, merger remnants become single HeWDs.
}
\label{m1m2}
\end{figure}

Also shown in Fig.~\ref{m1m2} shows the combinations of WD and MS mass at the point at which a MS star fills its Roche lobe that are realised at any time in our set of $10^7$ systems.
All systems in the region shown are unstable when CAML is included.
Systems in the region between the solid and dashed lines merge and form single HeWDs.
The decrease in density below $\MMS \la 0.35\,\Msolar$ is likely due to the fact that such MS stars are treated as fully convective in \textsc{bse} so that magnetic braking is switched off; without this angular momentum-loss mechanism to bring the MS to fill its Roche lobe, fewer WD+MS systems with $\MMS \la 0.35\,\Msolar$ merge.
These results are processed to find the properties of a model population with the chosen age and star formation rate history.

Assuming $\eta=0.5$ and that recently formed single HeWDs are those formed from mergers later than $11\,\rmn{Gyr}$, we find the Galactic formation rate of low-mass WDs to be about $8.7 \times 10^{-3}\,\rmn{yr}^{-1}$.
This value is nearly $2$ per cent of the total WD formation rate of $0.5\pm 0.125\,\rmn{yr}^{-1}$ inferred by \citet{Liebert2005} from observations,
when the volume of the Galaxy is taken to be $5\times10^{11}\,\rmn{pc}^3$.
This study concluded that approximately $10$ per cent of all WDs in the solar neighbourhood are low-mass WDs, of which less than $20$--$30$ per cent are single WDs.
Thus, less than about $2$--$3$ per cent of all WDs are single low-mass WDs, in agreement with out calculation.

\begin{figure}
\centering
\includegraphics[width=84mm]{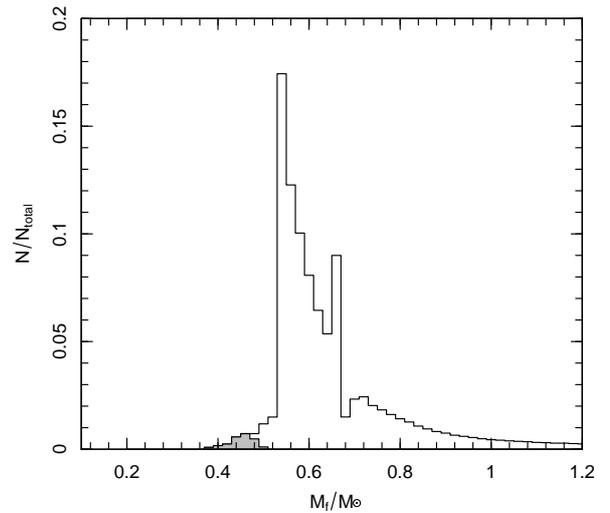}
\caption[]{Mass distribution of single WDs in our model population.
The mass distribution of single HeWDs is shaded grey.
All single WDs with masses $\le 0.48\,\Msolar$ are HeWDs formed after HeWD+MS mergers.}
\label{md}
\end{figure}

Fig.~\ref{md} shows the calculated mass distribution of all single WDs.
The single WDs are made up of
(1) COWDs and ONeMgWDs from very wide binaries, i.e., single stars, $94.9$ per cent of all single WDs;
(2) COWDs from HeWD+MS mergers, $2.7$ per cent of all single WDs; and
(3) HeWDs from HeWD+MS mergers, $2.4$ per cent of all single WDs (the shaded distribution).
In our model, all single WDs with masses less than $0.48\,\Msolar$ are HeWDs, and are formed from HeWD+MS mergers.

\begin{figure}
\centering
\includegraphics[width=84mm]{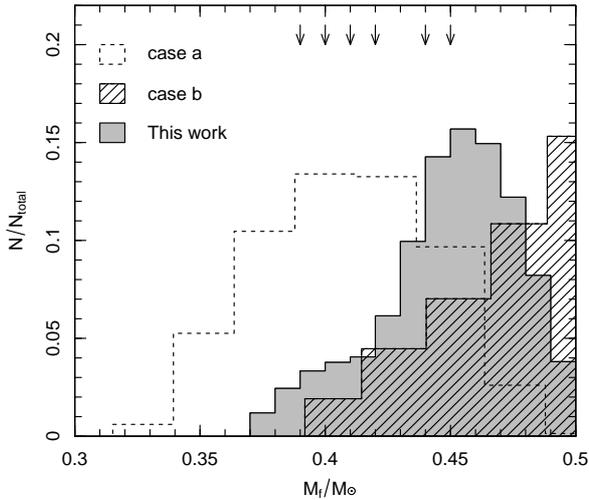}
\caption[]{Mass distribution of single low-mass WDs formed after HeWD+MS mergers in our model population (shaded grey region), and in cases a (dashed line) and b (hatched region) of \citet{Zorotovic2017}.
The arrows indicate the masses of single low-mass WDs observed by \citet{Liebert2005}.
A typical error of their masses are $0.02\,\Msolar$.}
\label{mass}
\end{figure}

Fig.~\ref{mass} shows the calculated mass distribution of single \emph{low-mass} HeWDs (shaded grey) and compares it to those of \citet{Zorotovic2017}; they performed similar calculations but approximated the post-merger evolution by assuming either (a) none of the MS star is accreted in the merger, or (b) the MS star is accreted in the merger and the remnant evolves at constant mass.
Our distribution spans a smaller range of masses than their work, between $0.37$ and $0.51\,\Msolar$, with a peak between $0.45$ and $0.46\,\Msolar$.
\citet{Zorotovic2017} found minimum masses $0.32\,\Msolar$ (case a) and $0.39\,\Msolar$ (case b).
The masses of single low-mass WDs observed by \citet{Liebert2005} and \citet{Brown11} are consistent with both our work and that of \citet{Zorotovic2017}.
The lowest mass WD observed by \citet{Liebert2005} was $0.39\,\Msolar$.
The typical uncertainties on their mass measurements were about $0.02\,\Msolar$.

\begin{figure}
\centering
\includegraphics[width=84mm]{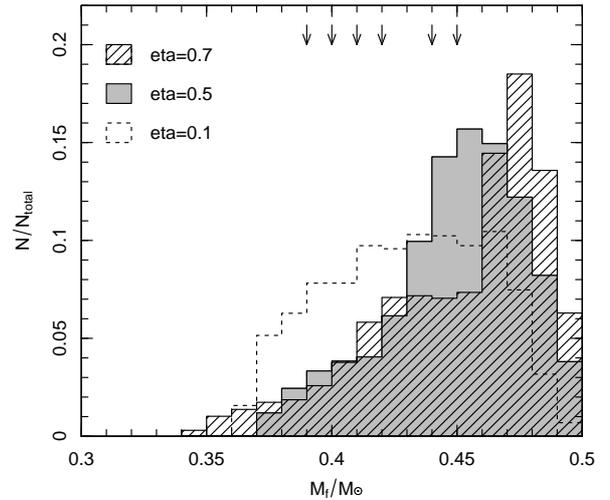}
\caption[]{Mass distribution of single low-mass WDs formed after HeWD+MS mergers in our model population when the Reimers' mass-loss parameter $\eta$ is varied.}
\label{eta}
\end{figure}

Calculations of the rate and mass distribution are affected by the assumed Reimers' mass-loss parameter $\eta$.
Assuming $\eta=0.5$, we found a Galactic formation rate of low-mass WDs about $8.7 \times 10^{-3}\,\rmn{yr}^{-1}$.
This rate changes to $1.3 \times 10^{-3}\,\rmn{yr}^{-1}$ with $\eta=0.1$ and $1.9 \times 10^{-2}\,\rmn{yr}^{-1}$ with $\eta=0.7$.
Fig.~\ref{eta} shows the effect of $\eta$ on the mass distribution of single low-mass WDs.
If $\eta=0.7$ then the minimum mass WD that can be formed by this channel is reduced to $0.34\,\Msolar$.
The minimum WD mass is reduced because the mass-loss rate from the merger remnant increases with $\eta$.
The final WD mass is dictated by the rate at which the core mass increases by shell burning and the rate at which the envelope mass decreases by mass-loss.
When the envelope mass becomes sufficiently small the star contracts away from the RGB to become a WD; the core mass at this point is the mass of the final WD to a good approximation.When $\eta=0.7$ the mass-loss rate from the merger remnant is larger than at lower $\eta$ so that the envelope mass becomes small when the core mass is smaller than it is in other cases.

\begin{figure}
\centering
\includegraphics[width=84mm]{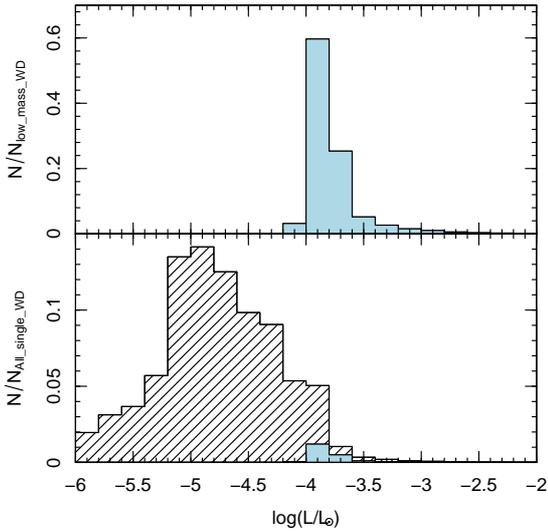}
\caption[]{Luminosity distribution of single WDs in our model population.
  Top panel: Luminosity distribution of single low-mass WDs formed after HeWD+MS mergers.
Bottom panel: Luminosity distribution of all single WDs with the distribution of single low-mass WDs formed after HeWD+MS mergers shaded blue.
The Reimers' mass-loss parameter $\eta=0.5$.}
\label{lumdis}
\end{figure}

Fig.~\ref{lumdis} shows the luminosity function of single WDs in our model population.
Comparison with an observed luminosity function could constrain our model but would require detailed consideration of selection effects.
The luminosity function depends on the formation age and cooling rates of WDs and is thus a potential discriminator between different formation channels to single low-mass WDs.

\section{Discussion and conclusion}
More than $90$ per cent of stars end their lives as WDs.
The initial mass of a star on the main sequence determines the mass of WD it will produce and the time it takes to become a WD.
Standard single star evolution does not account for single low-mass WDs, those with masses less than $0.5\,\Msolar$, because they would only form from initially low-mass stars after a time longer than the Hubble time.
The observed single low-mass WDs thus probably formed from binary star systems.
We have investigated the possibility that they are formed after HeWD+MS mergers.
Supplementing our models of the evolution of the remnants of such mergers, we combined our results with population synthesis.
We confirmed that the HeWD+MS merger channel to the formation of single low-mass WDs is reasonable and calculated the mass distribution of WDs formed in this way.
These calculations show that HeWD+MS mergers lead to the formation of single low-mass WDs in the Galaxy at a rate of about $8.7 \times 10^{-3}\,{\rm yr}^{-1}$, about
$2$ per cent of the total WD formation rate according to recent determinations.
Predicted masses range between $0.37$ and $0.51\,\Msolar$ and peak between $0.45$ and $0.46\,\Msolar$.
Our experiments on varying the post-merger mass-loss parameter $\eta$ suggest that an observed single low-mass WD with a mass less than about $0.34\,\Msolar$ would be difficult to explain as a HeWD+MS merger remnant.
Furthermore, we predict that a small number of HeWDs with masses larger than $0.5\,\Msolar$ are formed by HeWD+MS mergers.
These stars would be difficult to distinguish from CO WDs, except perhaps by asteroseismology.

\section*{Acknowledgements}
This work is supported by the grants 11703001, 10933002 and 11273007 from the National Natural Science Foundation of China,
the Joint Research Fund in Astronomy (U1631236) under cooperative agreement between the National Natural Science Foundation of China (NSFC) and Chinese Academy of Sciences (CAS), the China Postdoctoral Science Foundation, and the Fundamental Research Funds for the Central Universities.
XZ thanks Dr Jie SU for the helpful discussions.
Armagh Observatory and Planetarium is supported by a grant from the Northern Ireland Department for Communities.
PDH and CSJ acknowledge support from the UK Science and Technology Facilities Council (STFC) Grant No. ST/M000834/1

\bibliographystyle{mnras}
\bibliography{mybib}

\appendix
\section{\textsc{mesa} inlist}\label{sec:mesa_inlist}
To evolve merger remnants with \textsc{mesa} the parameters that differ from the defaults are as follows:
{\small
\begin{verbatim}

&star_job
   change_net = .true.
   new_net_name = 'agb.net'
/

&controls
   use_Type2_opacities = .true.
   initial_z = 0.02
   Zbase = 0.02

   mixing_length_alpha = 1.9179

   which_atm_option = 'Eddington_grey'

   cool_wind_RGB_scheme = 'Reimers'
   Reimers_scaling_factor = 0.5
   cool_wind_AGB_scheme = 'Blocker'
   Blocker_scaling_factor = 0.5
   RGB_to_AGB_wind_switch = 1d-4

   varcontrol_target = 1d-3
   mesh_delta_coeff = 2

   do_element_diffusion = .true.
   diffusion_dt_limit = 3.15d7
   diffusion_min_dq_at_surface = 1d-12
   surface_avg_abundance_dq = 1d-12
/
\end{verbatim}
}

\label{lastpage}
\end{document}